\documentclass[11pt]{article}
\usepackage{amsmath} 
\usepackage{amssymb}
\usepackage{verbatim} 
\usepackage{bm} 
\usepackage[colorlinks=true,citecolor=blue,linkcolor=black]{hyperref}
\usepackage{authblk} 
\usepackage{graphicx}

\usepackage{sectsty} 
\sectionfont{\large}
\subsectionfont{\normalsize}
\subsubsectionfont{\normalsize}
\numberwithin{equation}{section} 

\def\k{{\bf k}}

\def\x{{\bf x}}

\def\grad{{\bm\nabla}}

\def\j{{\bf j}}
\def\Eq#1{Eq.~(\ref{#1})}
\def\qmax{q_{\rm max}}

\title{ Fluctuation bounds on charge and heat diffusion
      }

\author{Pavel Kovtun}
\affil{
    Department of Physics and Astronomy, University of Victoria, Victoria, BC, V8P 5C2, Canada \vspace{-4ex}}

\date{}

\begin{document}

\maketitle

\begin{abstract}
\noindent
We study thermal fluctuation corrections to charge and heat conductivity in systems with locally conserved energy and charge, but without locally conserved momentum. Thermal fluctuations may naturally lead to a lower bound on diffusion constants for thermoelectric transport, and need to be taken into account when discussing potential bounds on transport coefficients. 

\end{abstract}

\section{Introduction}
In an interacting many-body system, the response to external sources at long distances is controlled by transport coefficients such as thermal conductivity, electrical conductivity, shear viscosity etc. A first-principles calculation of these transport coefficients is not an easy problem, especially when quantum fluctuations are strong, and it is of interest to look for general model-independent constraints on transport coefficients. One example of a powerful constraint is provided by Onsager relations which follow from time-reversal invariance. For transport coefficients in fluids, another set of constraints may be found from a local version of the second law of thermodynamics. 
There has also been some interest in constraints on transport coefficients which take the form of a lower bound. One often discussed example is a putative lower bound on the shear viscosity~\cite{Kovtun:2004de} which suggests that quantum fluctuations prevent the existence of perfect fluids in nature. The two fluids that come closest to the viscosity bound are cold atomic gases and the quark-gluon plasma, see~\cite{Schafer:2009dj} for a review. In solid-state physics, the Mott-Ioffe-Regel conductivity bound and its violations have been discussed for many years, see~\cite{RevModPhys.75.1085} for a review. 
Recently, Ref.~\cite{Hartnoll:2014lpa} argued that the diffusion constants for thermoelectric transport may be subject to a universal quantum lower bound. The bound of Ref.~\cite{Hartnoll:2014lpa} is imprecise, and needs better understanding. 
Here we point out that thermal fluctuations need to be taken into account when discussing a potential lower bound. A similar argument for the shear viscosity was put forward in~\cite{Kovtun:2011np}. The qualitative lesson is the same:  small values of transport coefficients, close to quantum lower bounds, may be subject to large thermal fluctuation corrections. How important these corrections are is determined by the thermodynamics of the system, and by the temperature dependence of the transport coefficients.

\section{Linear response}
Consider a system which conserves energy and charge, whose local conservation laws are
\begin{equation}
  \partial_t \epsilon + \grad{\cdot}\j_\epsilon = 0\,,\ \ \ \ 
  \partial_t n + \grad{\cdot}\j_n=0\,,
\label{eq:en}
\end{equation}
where $\epsilon$ is the energy density $n$ is the charge density, and $\j_\epsilon$, $\j_n$ are the corresponding spatial currents. In local thermal equilibrium in the grand canonical ensemble, $\epsilon=\epsilon(T,\mu)$, and $n=n(T,\mu)$, where $T$ is the temperature and $\mu$ the chemical potential corresponding to the conserved charge. In the hydrodynamic approximation, the spatial currents are expressed in terms of $T$, $\mu$ and their derivatives,
\begin{equation}
  \j_\epsilon = -\Pi_{11} \grad T - \Pi_{12} \grad\mu + \dots\,,\ \ \ \ 
  \j_n = -\Pi_{21}\grad T - \Pi_{22}\grad\mu + \dots\,,
\label{eq:jj}
\end{equation}
where $\Pi_{AB}$ are transport coefficients which depend on $T$ and $\mu$. The coefficient $\Pi_{22}$ is  the usual electrical conductivity $\sigma$. The dots denote higher-order terms in the derivative expansion. Supplementing \Eq{eq:en} with local momentum conservation would make the system behave as a normal fluid at long distances. Here we are interested in systems where momentum is not locally conserved, so that there are no other conserved densities besides $\epsilon$ and $n$ which are relevant in the hydrodynamic limit. In a physical system, momentum non-conservation may be due to lattice umklapp scattering, or due to impurities. The transport of charge and heat at long distances is then controlled by the transport coefficients in \Eq{eq:jj}. 

For linear fluctuations in thermal equilibrium, one takes the coefficients $\Pi_{AB}$ in \Eq{eq:jj} to be constant. The conserved densities $\varphi_A\equiv(\delta\epsilon,\delta n)$ are related to the corresponding sources $\lambda_A\equiv(\delta T /T, \delta\mu - \frac{\mu}{T}\delta T )$ by the equilibrium susceptibility matrix
\begin{equation}
\label{eq:chi}
  \chi = \begin{pmatrix}
  T \left(\frac{\partial\epsilon}{\partial T}\right)_{\mu/T} & 
  \left(\frac{\partial\epsilon}{\partial\mu}\right)_T\\[8pt]
  T\left(\frac{\partial n}{\partial T}\right)_{\mu/T} & \left(\frac{\partial n}{\partial\mu}\right)_T
  \end{pmatrix}
\end{equation}
which is symmetric and positive-definite. 
These properties of $\chi$ follow from the definition of the thermodynamic derivatives in the grand canonical ensemble: $\chi_{11}$ is proportional to the mean square energy fluctuation $\langle E^2\rangle_{\rm conn}$, while $\chi_{33}$ is proportional to the mean square charge fluctuation $\langle N^2 \rangle_{\rm conn}$ (the subscript denotes the ``connected'' average $\langle E^2\rangle - \langle E\rangle^2$ etc). The determinant of $\chi$ is proportional to $\langle E^2\rangle_{\rm conn} \langle N^2\rangle_{\rm conn} - \langle EN \rangle_{\rm conn}^2$ which is non-negative by the Schwarz inequality. We will assume a stable equilibrium state in which $\det \chi$ is strictly positive and the diffusion coefficients (see below) are finite.

Fluctuations of the conserved densities are related to $\psi_A\equiv(\delta T,\delta\mu)$ by the matrix of thermodynamic derivatives, $\varphi_A=X_{AB}\psi_B$. The conservation equations~(\ref{eq:en}) combined with the constitutive relations (\ref{eq:jj}) can then be written as
\begin{equation}
  \partial_t \varphi_A = D_{AB} \grad^2 \varphi_B\,,
\label{eq:D1}
\end{equation}
where the matrix of diffusion constants is $D=\Pi X^{-1}$, with
\begin{equation}
  \Pi \equiv \begin{pmatrix} \Pi_{11} & \Pi_{12} \\ \Pi_{21} & \Pi_{22} \end{pmatrix}\,,\ \ \ \ 
  X \equiv \begin{pmatrix}
  \frac{\partial\epsilon}{\partial T} & 
  \frac{\partial\epsilon}{\partial\mu}\\[8pt]
  \frac{\partial n}{\partial T} & \frac{\partial n}{\partial\mu}
  \end{pmatrix}\,.
\label{eq:PiX}
\end{equation}
The partial derivatives are at fixed $T$ or $\mu$, unless otherwise specified, and $T \det X = \det\chi >0$.
Following the standard linear response theory~\cite{KM}, the retarded functions of energy density and charge density $G^R_{AB}=\langle\varphi_A \varphi_B\rangle$ are given by
\begin{equation}
  G^R(\omega,\k) = ({\bf 1} + i\omega K^{-1})\chi\,,
\label{eq:GR}
\end{equation}
where $K\equiv -i\omega{\bf 1} + D\k^2$, suppressing the matrix indices. Time-reversal invariance requires that $G^R_{\epsilon n}(\omega,\k) = G^R_{n\epsilon}(\omega,-\k)$, which gives the Onsager relation 
\begin{equation}
  \Pi_{12}=T\,\Pi_{21} + \mu\, \Pi_{22}\,.
\label{eq:OR}
\end{equation}
Note that the Onsager relation does not imply that the matrix of diffusion constants must be symmetric.

The retarded functions have poles at $\omega=-iD_{1,2}\k^2$, where $D_{1,2}$ are the eigenvalues of the diffusion matrix $D$. The eigenvalues are not necessarily real, but may come as a complex conjugate pair, corresponding to the two solutions of the characteristic equation
$$
  (D_{1,2})^2 - D_{1,2}\, {\rm tr}(\Pi X^{-1}) + \det(\Pi X^{-1}) =0\,,
$$
so that $D_1 + D_2 = {\rm tr}(\Pi X^{-1})$, and $D_1 D_2 = \det(\Pi X^{-1})$.
The poles of the retarded function must be in the lower complex half-plane, hence one must have ${\rm Re}(D_{1,2}) >0$. This amounts to $D_1 + D_2 >0$ and $D_1 D_2 >0$, 
and implies the constraint on the transport coefficients
\begin{align}
\label{eq:cc}
  {\rm tr}\, (\Pi X^{-1}) > 0\,,\ \ \ \ \det \Pi >0\,.
\end{align}
If one demands that ${\rm Im}\,D_{1,2} = 0$, the characteristic equation gives the constraint
\begin{equation}
    {\rm tr}\, (\Pi X^{-1})  \geqslant 2\left( \frac{\det \Pi}{\det X}\right)^{1/2}\,.
\end{equation}

Taking the limit $\k\to0$ in the retarded functions gives rise to Kubo formulas for $\Pi_{AB}$ in terms of imaginary parts of $G^R_{\epsilon\epsilon}$, $G^R_{\epsilon n}$, and $G^R_{nn}$.
The imaginary part of the diagonal functions must have a definite sign, which implies 
\begin{equation}
  T\,\Pi_{11} + \mu\,\Pi_{12} \geqslant0\,,\ \ \ \ \Pi_{22}\geqslant0\,.
\end{equation}
Note that having vanishing electrical conductivity $\Pi_{22}$ and non-zero thermoelectric conductivity $\Pi_{12}$ is not allowed: in this case the Onsager relation~(\ref{eq:OR}) implies $\det \Pi <0$, which violates the causality constraint~(\ref{eq:cc}). Thus charge and temperature fluctuations must decouple as $\Pi_{22}$ goes to zero.

Current conservation combined with rotation invariance allows one to write the Kubo formulas in terms of the symmetrized functions, $G^S = 2T/\omega\; {\rm Im}\, G^R$ (assuming $\omega\ll T$), so that
\begin{subequations}
\label{eq:Kubo-S}
\begin{align}
   T\,\Pi_{11} + \mu\,\Pi_{12} & = \frac{1}{2Td}\, G^S_{j_{\epsilon,i}\, j_{\epsilon,i}}(\omega,\k{\to}0)\,,\\[5pt]
   \Pi_{12} & = \frac{1}{2Td}\, G^S_{j_{\epsilon,i}\, j_{n,i}}(\omega,\k{\to}0)\,,\\[5pt]
   \Pi_{22} & = \frac{1}{2Td}\, G^S_{j_{n,i}\, j_{n,i}}(\omega,\k{\to}0)\,.
\end{align}
\end{subequations}
Here $d$ is the number of spatial dimensions, the spatial indices on the currents are summed over, and $T$ and $\mu$ are the equilibrium temperature and chemical potential. We will use \Eq{eq:Kubo-S} as the definition of transport coefficients.

\section{Fluctuation corrections}
We now go one step beyond linear response. We would like to take into account the terms which are quadratic in $\delta T$, $\delta\mu$ in the right-hand side of~\Eq{eq:jj}. In the hydrodynamic limit $\k\to0$, the most important non-linear terms arise from the $T$ and $\mu$ dependence of the transport coefficients in~\Eq{eq:jj}. We can write~\Eq{eq:jj} more compactly as
$$
  \j_A = -\Pi_{AB}(\psi)\grad\psi_B + \dots\,,
$$
where $\Pi$ is given by \Eq{eq:PiX}, $\psi_A=(\delta T,\delta\mu)$ as before, and the currents are $\j_A=(\j_\epsilon, \j_n)$. Expanding to quadratic order in small fluctuations near equilibrium gives
\begin{equation}
  \j_A = - \Pi_{AB}\grad\psi_B - \Pi_{AB,C}\, \psi_C \grad\psi_B + \dots\,,
\label{eq:j2}
\end{equation}
where the leading coefficients $\Pi_{AB}$ are $\psi$-independent. A priori, ${\Pi}_{AB,C}$ are independent transport coefficients of the non-linear response. Neglecting fluctuations, they can be identified with $\partial\Pi_{AB}/\partial\psi_C$, evaluated in equilibrium, and so are determined by the $T$ and $\mu$ dependence of the linear-response transport coefficients. Namely, ${\Pi}_{AB,1} = \partial \Pi_{AB}/\partial T$ and ${\Pi}_{AB,2} = \partial \Pi_{AB}/\partial \mu$ in equilibrium. The eight coefficients ${\Pi}_{AB,C}$ have to satisfy two constraints which follow from the Onsager relation~(\ref{eq:OR}).

The quadratic terms in \Eq{eq:j2} induce the following fluctuation correction to the symmetrized function of the dot product of $\j_A$ and $\j_B$: 
$$
  \delta G^S_{AB}(t,\x) = {\Pi}_{AC,D} {\Pi}_{BE,F}\, 
  \langle \psi_D(t,\x) \grad \psi_C(t,\x)\; \psi_F(0) \grad \psi_E(0) \rangle^S\,.
$$
Factorizing the average gives the following one-loop expression
\begin{align}
  \delta G^S_{AB}(\omega,\k) = {\Pi}_{AC,D} {\Pi}_{BE,F} \int\frac{d\omega'}{2\pi} \frac{d^dk'}{(2\pi)^d}\, \Big( 
  & \k'^2 \Delta^S_{CE}(\omega',\k') \Delta^S_{DF}(\omega{-}\omega',\k{-}\k') \nonumber\\
  + & \k'{\cdot}(\k{-}\k') \Delta^S_{DE}(\omega',\k') \Delta^S_{CF}(\omega{-}\omega',\k{-}\k')
  \Big)\,.
\label{eq:deltaG}
\end{align}
Here the propagators are $\Delta^S = X^{-1} G^S (X^{-1})^T$, and $G^S_{AB} = \langle \varphi_A \varphi_B \rangle^S$ is the symmetrized two-point function of the conserved densities. One way to view this correction is to add Gaussian short-distance noise to the spatial currents $\j_n$ and $\j_\epsilon$, whose strength is determined by the fluctuation-dissipation theorem. This description may be readily converted to field theory, see e.g.~\cite{Kovtun:2012rj}. The free theory of linear response has a scaling symmetry, with time and space scaling as $t\sim x^2$. The coefficients corresponding to the non-linear terms in the constitutive relations~(\ref{eq:j2}) will scale proportional to inverse powers of momentum. The corresponding scale is the cutoff of the effective theory. \Eq{eq:deltaG} is then the one-loop correction due to the leading irrelevant operator in the effective theory.

To find the correction to transport coefficients, we need to evaluate~(\ref{eq:deltaG}) at $\k{=}0$. It is convenient to diagonalize $G^S$ by passing to the basis of eigenvectors of the diffusion matrix, at which point the calculation is straightforward. The integral diverges at large momenta, and we regulate it with a cutoff at~$q=\qmax$. The final answer is
\begin{subequations}
\label{eq:deltaGAB}
\begin{align}
 &  \delta G^S_{11} = \frac{T}{\det X} \left( \Pi_{11,2} - \Pi_{12,1} \right)^2 Y_{12}\,,\\[5pt]
 &  \delta G^S_{12} = \delta G^S_{21} = 
    \frac{T}{\det X} \left( \Pi_{11,2} - \Pi_{12,1} \right) \left( \Pi_{21,2} - \Pi_{22,1} \right) Y_{12}\,,\\[5pt]
 &  \delta G^S_{22} = \frac{T}{\det X} \left( \Pi_{21,2} - \Pi_{22,1} \right)^2 Y_{12}\,,
\end{align}
\end{subequations}
where
\begin{subequations}
\label{eq:Y12}
\begin{align}
  & Y_{12} = (2T)^2 \left[ \frac{1}{12\pi^2} \frac{\qmax^3}{D_1{+}D_2}
                        -\frac{1}{2\pi} \frac{|\omega|^{3/2}}{[2(D_1{+}D_2)]^{5/2}}\right]\,,\ \ \ \ d=3\,, \\[5pt]
  & Y_{12} = (2T)^2 \left[ \frac{1}{8\pi} \frac{\qmax^2}{D_1{+}D_2}
                        -\frac{|\omega|}{16(D_1{+}D_2)^2} \right]\,,\ \ \ \ d=2\,.
\end{align}
\end{subequations}
Following Kubo formulas (\ref{eq:Kubo-S}), expressions (\ref{eq:deltaGAB}) divided by $2Td$ can be interpreted as thermal fluctuation corrections to $T\,\Pi_{11}+\mu\,\Pi_{12}$, $\Pi_{12}$, and $\Pi_{22}$, respectively. We pause to make several comments.

The non-trivial correction to transport coefficients arises because of the coupling between charge density and energy density fluctuations, which manifests itself in a non-trivial index structure in~\Eq{eq:deltaG}. If charge density and energy density fluctuations were decoupled, the two terms in~\Eq{eq:deltaG} would cancel at $\k=0$.

Fluctuation corrections produce non-trivial contributions to transport. For example, if the thermoelectric coefficient $\Pi_{21}$ naively vanishes in linear response, a non-zero value for $\Pi_{21}$ will be induced by thermal fluctuations due to non-vanishing $\partial\Pi_{22}/\partial T$ and $\partial\Pi_{11}/\partial\mu$. The physical measured transport coefficients take into account all contributions, including those in~\Eq{eq:deltaGAB}.

The $|\omega|^{d/2}$ terms in \Eq{eq:Y12} correspond to $t^{-(d+2)/2}$ falloff of the correlation function in real time. This is well known, see e.g.~\cite{Arnold:1997gh, PhysRevB.73.035113}, and has the same physics as the long-time tails in fluids discovered by Alder and Wainwright~\cite{PhysRevA.1.18}. The coefficients of the $|\omega|^{d/2}$ terms in~\Eq{eq:Y12} are cutoff-independent. 

The momentum cutoff has to be taken sufficiently large, $\omega\ll (D_1{+}D_2)\qmax^2$ to ensure the validity of the hydrodynamic description~(\ref{eq:jj}) at long distances. The momentum cutoff also has to be sufficiently small so that the higher-derivative terms in (\ref{eq:jj}) (which we ignored) do not contribute. Physically, $1/\qmax$ is the length scale at which the hydrodynamic description (\ref{eq:jj}) breaks down.

The corrections to the fundamental transport coefficients $\Pi_{AB}$ can be readily found from (\ref{eq:deltaGAB}) and (\ref{eq:Kubo-S}), and can be used to find the corrections to the diffusion constants~$D_{1,2}$. However, as there are three independent transport coefficients $\Pi_{AB}$, and only two diffusion constants $D_{1,2}$, there is no reason to expect that the fluctuation corrections to $D_1$ and $D_2$ are expressible in terms of $D_1$ and $D_2$. This does happen to be the case for the sum though, and the fluctuation correction to $D_1{+}D_2$ takes a particularly simple form, 
$$
  \delta (D_1 {+} D_2) = \frac{1}{2Td} \,{\rm tr} \left( \chi^{-1} \delta G^S\right)\,,
$$
where $\chi$ is the susceptibility matrix in (\ref{eq:chi}), and $\delta G^S$ is given by (\ref{eq:deltaGAB}). 
Schematically, the correction (\ref{eq:deltaGAB}) is $\delta G^S_{11}=a^2$, $\delta G^S_{12}=\delta G^S_{21}=ab$, $\delta G^S_{22}=b^2$, and ${\rm tr} \left( \chi^{-1} \delta G^S\right) = z^T \chi z/\det\chi$, where $z\equiv(b,-a)^T$. Positive definiteness of $\chi$ then ensures that the correction to $D_1{+}D_2$ is positive.

The correction to $D_1{+}D_2$ is inversely proportional to $D_1{+}D_2$. Explicitly, if $D_1^b$ and $D_2^b$ are the naive ``bare'' values, the full diffusion constants are
\begin{equation}
\label{eq:DD}
  (D_1 {+} D_2) = (D_1^b {+} D_2^b) + \frac{\qmax^d C_d }{(D_1^b {+} D_2^b)} \,,
\end{equation}
where $C_d$ is determined by the derivatives of transport coefficients with respect to $T$ and $\mu$,
\begin{align}
\label{eq:Cd}
  C_d =  \frac{ c_d \, T}{\left(\frac{\partial\epsilon}{\partial T} \frac{\partial n}{\partial\mu} - \frac{\partial\epsilon}{\partial\mu} \frac{\partial n}{\partial T}\right)^2} & \left[ 
  T\frac{\partial\epsilon}{\partial T} \left(\Pi_{22,1} {-} \Pi_{21,2} \right)^2 
  + \frac{\partial n}{\partial\mu}\left(\Pi_{11,2}{-}\Pi_{12,1} + \mu(\Pi_{22,1}{-}\Pi_{21,2})\right)^2
  \right. \nonumber\\
  & \left. +\,  T\frac{\partial n}{\partial T} \left(\Pi_{22,1} {-} \Pi_{21,2}\right)
    \left(2(\Pi_{11,2} {-} \Pi_{12,1}) +\mu(\Pi_{22,1} {-} \Pi_{21,2}) \right) \right]\,.
\end{align}
The numerical coefficients are $c_3 = 1/18\pi^2$ and $c_2 = 1/8\pi$.

\section{Discussion}

\begin{figure}
\begin{minipage}{.5\textwidth}
  \centering
  \includegraphics[width=0.9\textwidth]{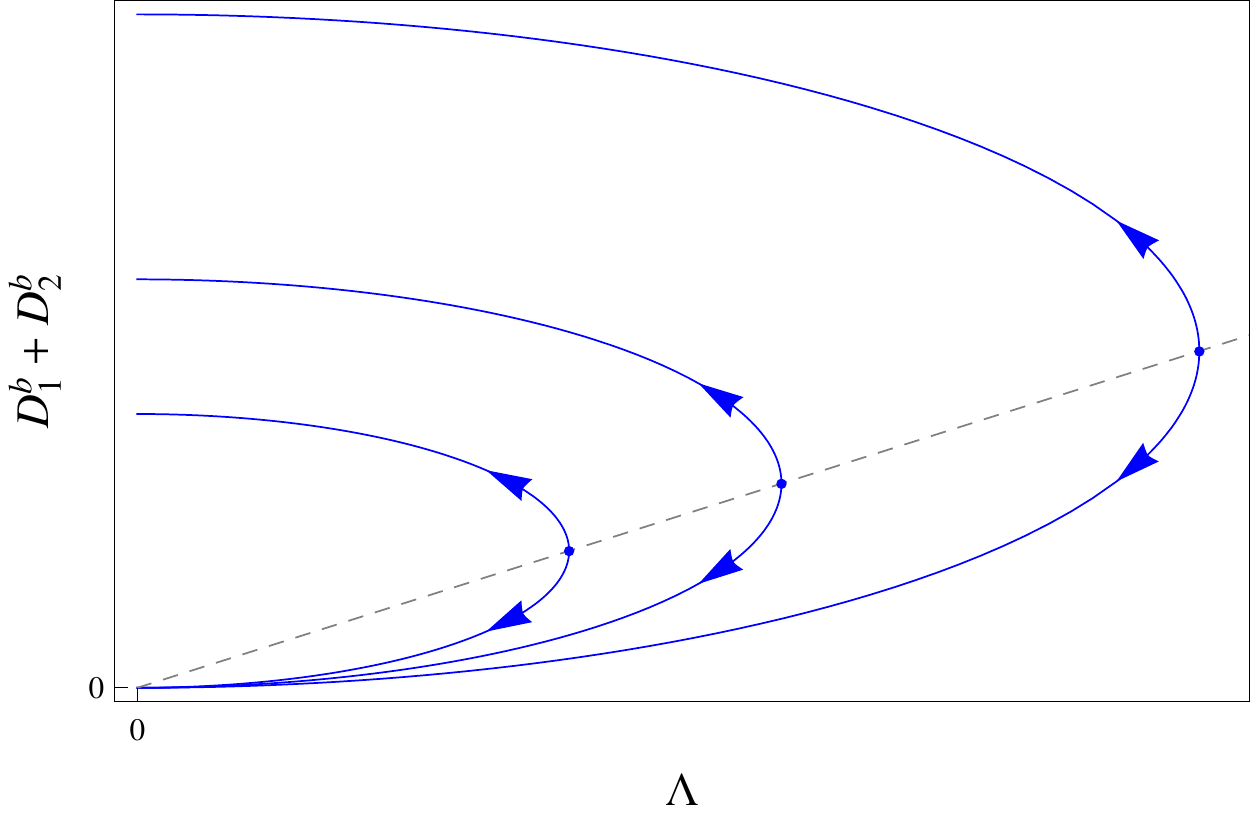}
\end{minipage}%
\begin{minipage}{.5\textwidth}
  \centering
  \includegraphics[width=0.9\textwidth]{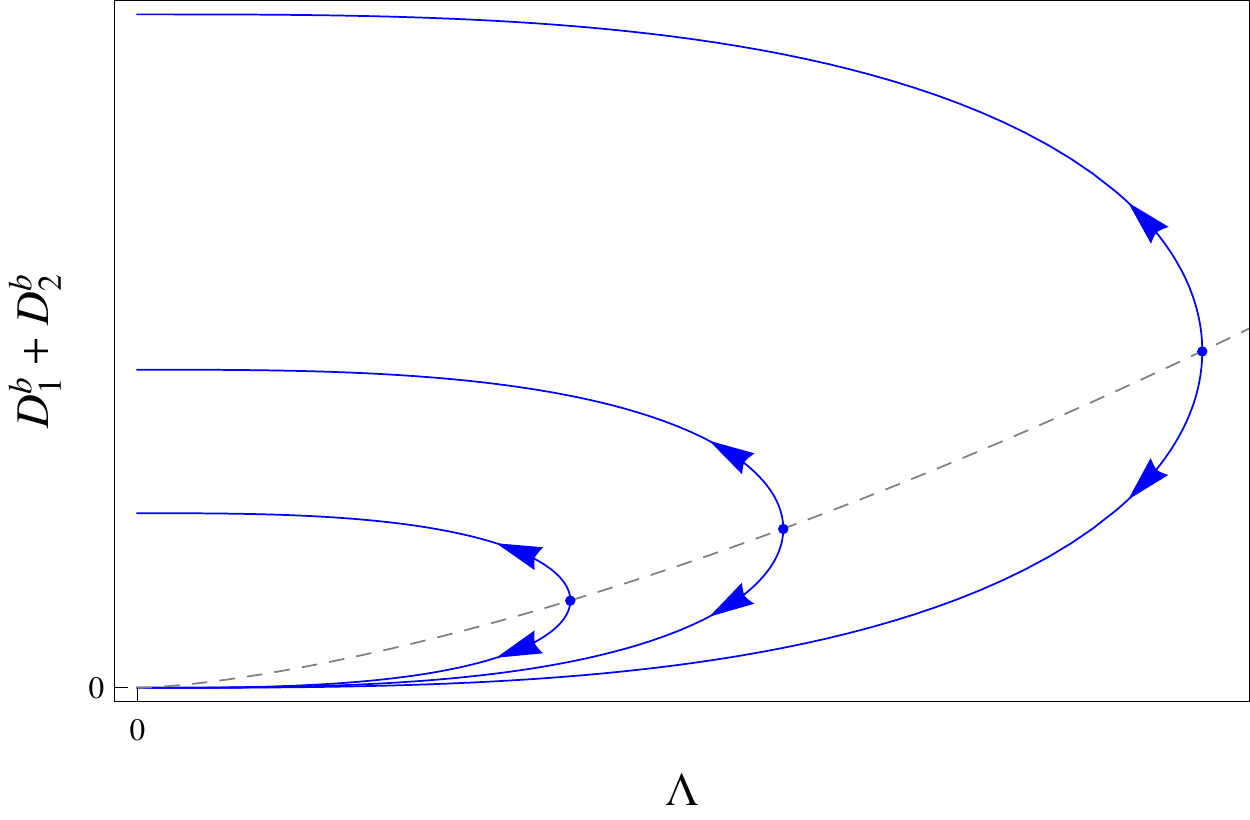}
\end{minipage}
\caption{
Flow diagram for \Eq{eq:rgeq} for $d=2$ (left) and $d=3$ (right), showing $(D_1^b{+}D_2^b)$ as a function of $\Lambda$. The dashed line separating the two flows is $(C_d \Lambda^d)^{1/2}$. Below the dashed line, the correction in \Eq{eq:DD} is large and the flow is not to be trusted. Arrows indicate the direction towards decreasing $\Lambda$. The physical value is at $\Lambda\to0$.
}
\label{fig:rg}
\end{figure}

The diffusion constants in \Eq{eq:DD} are written as a sum of two terms: the ``bare'' contribution from the modes at the cutoff scale, and the correction from the hydrodynamic modes below the cutoff. 
One way to look at the corrections in (\ref{eq:DD}) is to take the cutoff as the sliding scale $\qmax=\Lambda$. The total (physical) $D_1{+}D_2$ must be cutoff independent,
\begin{equation}
\label{eq:rgeq}
 \frac{d}{d\Lambda} \left( D_1 + D_2 \right) = 0\,,
\end{equation}
which gives $(D_1^b{+}D_2^b)$ as a function of $\Lambda$. 
For the leading correction to $(D_1{+}D_2)$, the coefficients $\Pi_{AB,C}$ are taken at their ``bare'' values, hence the coefficient $C_d$ is $\Lambda$-independent.
The solutions to Eq.~(\ref{eq:rgeq}), (\ref{eq:DD}) are sketched in Figure~\ref{fig:rg}. The dashed line is the set of points where the two terms in the right-hand side of~(\ref{eq:DD}) are equal in magnitude. Below the dashed line, the correction to $D_1^b{+}D_2^b$ is greater than the leading term, and the picture is not to be trusted. In the region above the dashed line where the correction is small and the calculation is reliable, one can see that $(D_1^b{+}D_2^b)$ grows as the momentum cutoff is lowered. 
This means that a lower bound on $(D_1^b{+}D_2^b)$ may be imposed consistently, thanks to the positivity of $C_d$: if one starts above the bound at the cutoff, one always ends up above the bound at long distances. 
From this point of view, a bound on $D_1{+}D_2$ appears to be more natural than a bound on $D_1$ and $D_2$ separately, as the latter are not necessarily real. 

Another way to look at the corrections is to keep the cutoff at a fixed physical scale. Then if the right-hand side of \Eq{eq:DD} has a minimum as a function of $D_1^b {+}D_2^b$, one may argue that the sum of the diffusion constants must be bounded from below. 
In order to place a lower bound on $D_1 {+} D_2$, one needs to know how the bare diffusion constants depend on the cutoff. 
This is because ultimately $D_{1,2}^b$ are determined by the physics at the cutoff scale, such as quasiparticle scattering, impurities etc. Different relations between $D_{1,2}^b$ and the cutoff will give rise to different bounds. 
For example, let us assume for simplicity that the thermoelectric transport is controlled by a single microscopic relaxation time scale~$\tau$. Then $(D_1^b{+}D_2^b)\qmax^2 \sim 1/\tau$, and Eq.~(\ref{eq:DD}) becomes $D_1 {+} D_2 = x + C_d\, \tau^{-d/2} x^{-(d+2)/2}$, for $x\equiv D_1^b{+}D_2^b$. This has a minimum as a function of $x$, and allows one to place a lower bound
\begin{subequations}
\label{eq:DD-bounds}
\begin{align}
  &  D_1 {+} D_2 \;\gtrsim\; \frac{C_d^{2/7}}{\tau^{3/7}}\,,\ \ \ \ d=3\,,\\[5pt]
  &  D_1 {+} D_2 \;\gtrsim\; \frac{C_d^{1/3}}{\tau^{1/3}}\,,\ \ \ \ d=2\,.
\end{align}
\end{subequations}
These are very non-trivial relations, with transport coefficients bounded by their thermodynamic derivatives, and by the microscopic relaxation time~$\tau$. 

The basic assumption of the paper was the hydrodynamic description of the near-equilibrium transport, supplemented by classical thermal fluctuations of the hydrodynamic modes. This describes macroscopic thermoelectric conduction and ignores localization effects. However, within the hydrodynamic framework, the exact microscopic origin of the transport coefficients does not matter in order to arrive at the correction (\ref{eq:DD}). The microscopic details show up in the relation between the diffusion constants and the momentum cutoff, and lead to bounds such as Eq.~(\ref{eq:DD-bounds}). The lower values saturating the bounds have to be taken with a certain grain of salt, as the correction becomes of the same order as the leading term at the minimum.

The qualitative form of the fluctuation corrections is the same regardless of whether the microscopic dynamics is weakly or strongly coupled. For example, in the ``holographic'' models, strongly interacting large-$N$ quantum field theories are described by classical gravity~\cite{Hartnoll:2009sz}. In such setups, fluctuation corrections will show up at a subleading order in the large-$N$ expansion. As a result, the holographic calculations that are based on purely classical gravity only capture the leading transport terms (e.g. the first term in Eq.~(\ref{eq:DD})), and are blind to the fluctuation effects. Quantum-gravitational holographic calculations, on the other hand, do see the fluctuation corrections~\cite{CaronHuot:2009iq}.

The main qualitative lesson is that small values of the diffusion constants at the cutoff scale will lead to large thermal fluctuation corrections: the smaller $D_{1,2}$ are, the larger the corrections are. As a result, using the linear response theory for metallic transport with very small diffusion constants may be qualitatively incorrect. Are the potential lower bounds on thermoelectric transport coefficients proposed in Ref.~\cite{Hartnoll:2014lpa} still large enough to justify using the linear response theory? To answer this question, one needs to know the value of the $C_d$ coefficient in Eq.~(\ref{eq:Cd}). This requires detailed knowledge of thermodynamics, as well as knowing the $T$ and $\mu$ dependence of all the transport coefficients~$\Pi_{AB}(T,\mu)$ which appear in Eq.~(\ref{eq:jj}).

The question about the applicability of the linear response theory is easier to answer for systems which conserve momentum. In that case, there is an extra hydrodynamic degree of freedom, the fluid velocity. The coefficients of the leading non-linear terms in the constitutive relations will be thermodynamic functions, and the right-hand side of the fluctuation bound will be determined by the equation of state alone, rather than by the derivatives of the transport coefficients. This has been applied to the quark-gluon plasma~\cite{Kovtun:2011np} and to the unitary Fermi gas~\cite{Chafin:2012eq, Romatschke:2012sf}, where lower bounds similar to Eq.~(\ref{eq:DD-bounds}) were placed on the shear viscosity. In those systems, detailed information about the equation of state is available, and it was found that the fluctuation correction to viscosity is indeed large if one takes the ratio of viscosity to entropy density close to the conjectured lower bound $\eta/s=\hbar/4\pi$. Whether something similar happens with diffusion constants in incoherent metals discussed in Ref.~\cite{Hartnoll:2014lpa} remains to be seen.

\subsection*{Acknowledgments}
I thank Adam Ritz for helpful conversations. This work was supported in part by NSERC of Canada.

\bibliographystyle{utphys}
\bibliography{dbound}

\end{document}